\documentstyle[prb,aps,epsfig,twocolumn]{revtex}
\begin{document}
\draft
\flushbottom
\twocolumn[
\hsize\textwidth\columnwidth\hsize\csname @twocolumnfalse\endcsname
\title{Boundary Conditions for Bulk and Edge States in Quantum Hall Systems}
\author{E.~Akkermans, J.~E.~Avron, R. Narevich\\ Department of Physics,
 Technion, 32000 Haifa, Israel \\and
R. Seiler\\Fachbereich Mathematik, TU-Berlin, Germany}
\date{\today}
\maketitle
\tightenlines
\widetext
\advance\leftskip by 57pt
\advance\rightskip by 57pt
\begin{abstract}
For two dimensional Schr\"odinger
Hamiltonians  we formulate boundary conditions that  split the Hilbert
space according to the chirality of the eigenstates on the boundary.
With magnetic fields, and in particular, for Quantum Hall systems,
this splitting
corresponds to  edge and bulk states.
Applications to the integer and fractional Hall effect and some open problems
are described.
\end{abstract}

\vskip 1cm
\pacs{PACS: 73.40.Hm, 71.70.Di, 73.23.-b, 02.60.Lj}

]

\narrowtext
\tightenlines


\setcounter{equation}{0}
\def\real{{\rm I\kern-.2em R}}
\def\complex{\kern.1em{\raise.47ex\hbox{
	    $\scriptscriptstyle |$}}\kern-.40em{\rm C}}
\def\integer{{\rm Z\kern-.32em Z}}



The theory  of the Quantum Hall Effect has been torn between several schools of
thought: one stresses the two dimensional bulk aspects of the interior
\cite{bulk}; another emphasizes the importance of the one dimensionality of the
edge \cite{edge} and   other points of view focus on the interplay between
bulk and edge \cite{halperin}. It is therefore remarkable that in spite
of this the notion of  bulk and edge of a quantum system
 has not been formulated as a sharp dichotomy even for idealized situations.
Classically,  there is such a  dichotomy  for  billiards
in magnetic fields: orbits
that lie in the interior rotate one way,
clockwise for positively charge particles,  while orbits that hit the
edge  make a skipping
orbit and rotate counter-clockwise \cite{pi}. Bulk and  edge are therefore
distinguished by the chirality relative to the boundary.  Our purpose here is to
formulate a corresponding dichotomy in quantum mechanics. As we shall explain
this can be achieved by imposing certain {\em chiral}  boundary
conditions for Schr\"odinger and Pauli operators.

The chiral boundary condition we introduce is sensitive to the   direction
of the (tangential) velocity  on the boundary. For (separable) quantum
billiards this enables us to split the one particle Hilbert space
into a direct sum of two orthogonal, {\em infinite dimensional} spaces with
positive and negative chirality on the boundary. In the presence of a
magnetic field,
this split gives a Hilbert space for edge states, ${\cal H}_e$,  and a
Hilbert space for bulk states,
${\cal H}_b$, such that the full Hilbert space is
${\cal H}={\cal H}_e\oplus {\cal H}_b$.
Subsequently we shall explain how chiral boundary
conditions are formulated for Schr\"odinger Hamiltonians which do not
necessarily correspond to  separable billiards, i.e.  Schr\"odinger
Hamiltonians
with background potential and electron-electron interactions.

 The chiral
boundary condition we introduce is a relative of a boundary condition
introduced by Atiyah, Patodi and Singer (APS)
 in their studies of Index theorems for Dirac operators with
boundaries \cite{aps}. However, the chiral boundary condition we shall
introduce differs from it in an important way, as we shall explain
below.

The splitting of the Hilbert space comes with a splitting of the quantum
billiard Hamiltonian and its spectrum  to a bulk piece and an edge piece. As we
shall see, it is a property of the chiral boundary conditions that the bulk
spectrum has  a ground state at precisely the energy of the lowest Landau level
in the infinite plane, and a degeneracy which is  the total flux through the
billiard, (corrected to an integer number of flux units by a boundary term).
The
bulk energy spectrum has a gap above the ground state, which for
separable
billiards, is  the gap between Landau levels in the infinite plane. Since this
gap survives in the thermodynamic limit of a billiard of infinite area, the bulk
ground state is guaranteed to be  incompressible in this sense.

In contrast, the edge spectrum,
in the thermodynamic limit of long boundary  is gapless.  In this
limit, the edge states have a well defined ''sound velocity", which reflects the
linearity of the dispersion relation at low energies. The sound velocity
$v$  is
\begin{equation}v/c= k {\lambda_c}\,/\ell_B,\label{compton}\end{equation}
where
$k$ is a dimensionless (nonuniversal) constant, c is the velocity of light ,
${\lambda_c}= {\hbar \over {mc}}$ is the Compton wavelength of the
electron and $\ell_B=\sqrt{\Phi_0 / B}$ is the magnetic length.
This sound velocity is very small in all
reasonable magnetic fields.

The splitting of the Hilbert space enables us to describe charge transport in
terms of spectral flow. In particular, (adiabatic) gauge transformations can
transfer  states between ${\cal H}_e$ and ${\cal H}_b$. For the semi-infinite
cylinder, such a spectral flow is described below. This
generalizes the Index theory of the Integer quantum Hall effect
\cite{ass} to systems with boundaries.

We start with the semi-infinite cylinder for which we shall illustrate the
chiral boundary condition.  The Landau
Hamiltonian with chiral boundary condition is separable and a complete
spectral analysis can be made.

Consider  the semi-infinite
cylinder, $M$, in $\real^3$, whose boundary $\partial M$ is a circle with a
circumference
$\ell$: $M=\{(x,y)\,|\, \ -\infty\le x \le 0, \  0\le y <\ell\  \}$. The
orientation of $M$ and the orientation of the  boundary, $\partial M$, are
linked by requiring that
traversing the boundary in the positive direction keeps
$M$ on the {\em left} hand side.

A constant magnetic field perpendicular to the
surface, of strength $B>0$ and with outward orientation acts on the surface.
We take the charge of the electron to be positive (sic!) so classical (bulk)
electrons in the interior of
$M$ rotate clockwise. In addition we assume that a flux tube carrying flux
$\phi$ threads the cylinder. We shall regard $\phi$ as a parameter, while $B$
is kept fixed throughout.
A gauge field describing the situation is
$A(\phi)=(0, Bx +\phi/\ell)$.   The velocity operator, in units $m=\hbar=e/c=1$,
is
$\left( v_x , v_y \right) =\left( -i \partial_x , - i
\partial_y - Bx - \phi/\ell \right)$. The classical energy associated to a
particle on a billiard is purely kinetic,
$E=v^2/2$. The corresponding quantum Hamiltonian  is the Landau Hamiltonian
given formally by the second order partial differential operator:
\begin{equation} 2H_L(\phi) = D^\dagger(\phi) {D}(\phi) +
B,\label{landau}\end{equation}  where
$D(\phi) =  iv_x - v_y(\phi,x)= \partial_x + (i\partial_y
+Bx+\phi/\ell)$.

For this to define a self-adjoint operator in the one particle Hilbert space we
need to specify boundary conditions on
$\partial M$.

The chiral boundary condition that we introduce requires
different things from the wave function on the boundary depending on  the
 tangential velocity,
$v_y(\phi,x)$ at  the boundary $x=0$.
Since $v_y(\phi,0)= -i\partial_y -
\phi/\ell$  commutes with $D$ we  separate variables, and describe the chiral
boundary conditions for the resulting ordinary differential operators  on the
half line $-\infty\le x
\le 0$, parameterized by
$m\in\integer$ and $\phi\in\real$:
\begin{equation}
\label{angularmomentum} 2H_m(\phi) = - \frac{d^2}{dx^2} + \left(\frac{2\pi m -
\phi}{\ell} - Bx \right)^2.
\end{equation}
Let
\begin{equation}
D_m(\phi) = \frac{d}{dx}- {\frac{2\pi m-\phi}{\ell}} +Bx.
\end{equation}
The chiral boundary condition requires:
\begin{eqnarray} D_mf_m\Big|_{x=0}=
 0,
\ &{\rm if}&
\ v_y(\phi,0)= \frac{2\pi m\ -\phi}{\ell} \le 0;\nonumber\\
(iv_x)\,f_m\Big\vert_{x=0}=
0, \ &{\rm if}&\
v_y(\phi,0)=  \frac{2\pi m\ -\phi}{\ell}> 0.
\label{spn}
\end{eqnarray}
Recall that a classical electron in the bulk rotate clockwise, and so its
velocity near the boundary {\em disagrees} with the orientation of the boundary.
For such an electron we require {\em spectral} boundary conditions,
$(D_mf)(0)=0$,
which are  $m$-dependent elastic boundary conditions (an interpolation between
Neumann and Dirichlet). A classical skipping orbit near the boundary moves in a
direction that agrees with the orientation of the boundary,  and  for
positive velocity on the boundary we impose Neumann boundary condition.  We
shall say more on the reasons for choosing spectral and Neumann  for the
different chiralities below.

Since both the differential operator, and the
boundary conditions are defined in terms of the  velocity,
gauge invariance is manifest. Moreover, it can be checked that the boundary
conditions  in Eq.~(\ref{spn}) define a self-adjoint  eigenvalue problem, which
we shall call the chiral Landau Hamiltonian. The spectrum and eigenfunctions
 can be described in terms of special functions \cite{tricomi}.

\begin{figure}
\begin{center}
  	\leavevmode
	\epsfig{file=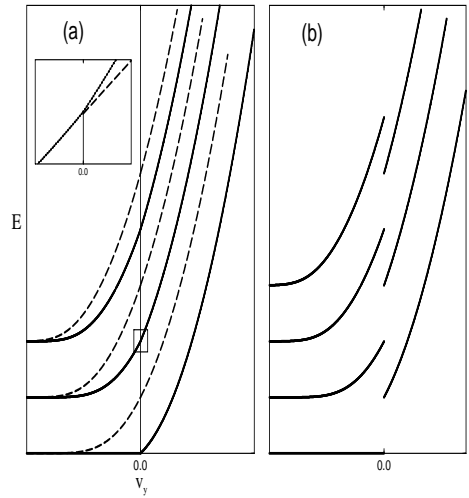,height=6cm,width=7cm,angle=270}
\end{center}
\caption{Spectrum of the Landau Hamiltonian.(a) with chiral (solid lines)
and Dirichlet (dashed lines) boundary conditions (inset- enlarged box,
showing a cusp between bulk and edge states), (b) with APS boundary
conditions.}
\label{fig:1}
\end{figure}

The bulk space ${\cal H}_b$  is defined by
\begin{equation}
{\cal H}_b = {\oplus\atop {\scriptstyle 2\pi m \le \phi}}\  e^{ 2\pi
imy/\ell } f_m (x) ,
\end{equation}
where $f_m$ are the eigenfunctions of the chiral Landau Hamiltonian that satisfy
spectral boundary condition.  ${\cal H}_e$,
the space of edge states, is the orthogonal complement.
The spectrum for
 the chiral Landau Hamiltonian is shown in Fig.~1.a as a collection
of curves plotted as functions of the velocity on the boundary. The bulk
spectrum is determined by the left part of the figure i.e. by negative
values of the velocity and
the edge spectrum by the right part (positive values).  The ground
state of the bulk
spectrum has energy $B/2$ which corresponds to the lowest Landau level in
the plane
(doubly infinite cylinder). Like it, it is infinitely
degenerate.
This turns out to be a  property of chiral boundary conditions that holds for a
large class of billiards: the ground state of the bulk spectrum has  energy
$B/2$ and the degeneracy is (an integer close to) the total flux through the
billiard.  The present case where the total flux is infinite is an example.  The
bulk ground state is separated by a gap
$B$ from the first excited bulk state.  For the excited bulk states the
situation is more complicated, and one general statement is that
the essential bulk spectrum, coincides with
the spectrum of the Landau Hamiltonian in the plane:
 that is, the bulk
spectrum differs from the Landau spectrum by at most a discrete set of
eigenvalues.

The edge spectrum, in contrast,
is, for any finite boundary length $\ell$, purely
discrete (the essential spectrum is empty). In the thermodynamic limit
$\ell\to\infty$  the
edge spectrum becomes gapless.  The slope of the curves describing the edge
spectrum
 give a linear dispersion with a finite sound velocity as
$v_y(\phi,0)\searrow 0$.
In particular, for the lowest edge branch one has, in
the limit $\ell\to\infty$,
a unique sound velocity for the chiral edge currents:
\begin{equation}
\frac{\partial E_0}{\partial v_y}\Big\vert_0=
\sqrt {B \over \pi}
\end{equation}
This fixes the  $k$ in Eq.~(\ref{compton})  in this case. It is worth
emphasizing the existence of the cusp between bulk states and the
corresponding edge branch as shown in Fig.1.a.

It is instructive to compare the spectral properties of the Chiral Landau
Hamiltonian with the  Dirichlet Landau model, where one replaces
Eq.~(\ref{spn}) by the requirement
$f_m(0)=0$ for all m. This too can be solved explicitly in terms of
special functions \cite{tricomi} and
the spectrum is shown in Fig.~1.a.  The corresponding curves unlike the
chiral case
 are analytic functions. This has some immediate implications:
First, there is no sharp line of divide between edge and bulk, second, there is
no natural sound velocity because the dispersion law is not linear at small
energies, and finally, there is no macroscopic degeneracy of the ground state
(or any other state).

The chiral boundary condition Eq.~(\ref{spn}) is a close relative
 of boundary
conditions introduced in [4].
APS boundary condition
replaces Eq.~(\ref{spn}) by
\begin{eqnarray} \left(\frac{d}{dx}- \frac{2\pi
m-\phi}{\ell}\right)f_m\Big|_{x=0} &=& 0
\quad {\rm if}
\ v_y(\phi,0) \le 0;\nonumber\\
f_m\Big|_{x=0} &=& 0 \quad {\rm if}\  \ v_y(\phi,0) > 0.
\label{aps}
\end{eqnarray}
That is, the Neumann piece for the edge states is replaced by
Dirichlet. Here too there is a sharp divide of the states according
to their chirality. But, in APS the putative edge states with the good chirality
are forced to have vanishing density near the boundary and tend to be pushed
away from the edge. These can not be bona fide edge states.  The APS Landau
Hamiltonian  can be solved explicitly for the problem at hand, and the spectrum
is shown in Fig. 1.b. The glaring difference with Fig.1.a is that
now the
energy curves are {\em discontinuous}. As we shall explain, this discontinuity
has undesirable features for studying spectral flows and transport in quantum
mechanics.

Consider now the spectral flow resulting from the increase of the
 threading flux $\phi$ by a unit
of quantum flux:
$\phi\to
\phi+2\pi$. By inspection of Fig.1 one sees that all states in
the diagrams move one notch to the left. In the chiral and APS cases which
have a clear
divide between chiralities we see that each branch of the good chirality
looses a state and each branch of the bad chirality gains one. In the
chiral case (Fig.1.a)  one
can follow continuously  each state as its chirality changes. In Fig.1.b
this is not the case.
Chiral boundary conditions therefore give a way of counting the charge being
transport  from bulk to edge. The same
spectral flow takes place for the Dirichlet spectrum except that here what is
edge and what is bulk is a vague notion which does not allow for counting 
the states that move from edge to bulk.
In the case of APS the notion of edge and bulk is sharp, but
 because of the
discontinuity of the curves in Fig.1.b there is no  way to
identify the flow of bulk to edge.

It is also instructive to examine how chiral boundary
conditions are related to Laughlin states. As we shall see, Laughlin states
for filling fraction $1/M$, $M$ an odd integer, are bulk
states with maximal density.

 To simplify the notation let us take a cylinder of
area
$2\pi$,
$M=\{(x,y)\,|\,
\ -1\le x \le 0, \  0\le y <2\pi\  \}.$
We shall take $\phi=0$ in what follows.
The Laughlin state of the (doubly infinite) cylinder for filling fraction
$1/M$, with $M$ odd is \cite{thouless}
\begin{equation}
\psi_L=\prod_{1\le j<k\le N}
\left({e^{{-z_j}}}-{e^{{-z_k}}}\right)^{M}\,\prod_{1\le k\le N}
e^{-Bx_k^2/2+m {z_k}}.
\end{equation}
Here $z=x+iy$ and $m \in\integer$. Fix a particle, say $z=z_1$. As a function
of
$z$, $\psi_L$ has the form
\begin{equation}
\left(A_1e^{-M(N-1)z}  + A_2 e^{-M(N-2)z}+\dots
\right)
e^{-Bx^2/2+m z}
\end{equation}
where $A_j$  are independent of $z$. The chiral boundary conditions for $z$ need
to be imposed on the two bounding circles at $x=0$ and $x=-1$ with opposite
orientations.  Since $\psi_L$ is in the kernel of $D$,
($D{\psi_L} =0$), the spectral
boundary
conditions are automatically satisfied. So, all that needs to be checked is that
the velocity on the two bounding circles is anti-chiral. That is:
\begin{equation}
m +B \ge M(N-j)\ge m,
\end{equation}
for all $1\le j \le N$. $j=N$ sets $m =0$, and $j=1$ sets an upper bound on
the number of electron that the Laughlin state may accommodate and still
satisfy the chiral boundary conditions :
$N\le 1 + B/M$.   Recall that  the area of the cylinder
is
$2\pi$, so that $B$ is the total flux in units of quantum flux. In the
(thermodynamic) limit of large
 flux  the maximal filling is $N/B\to 1/M$, which is what Laughlin plasma
argument gives \cite{laughlin}.

The case of other separable billiards, such as a circular disc  can be treated
in a similar way. The new feature that arises for separable billiard of finite
area is that there are interesting index theorems for the degeneracy of the
chiral bulk ground state.  These issues will be described elsewhere \cite{an}.

We now turn to the
description of the chiral boundary conditions for more general
Schr\"odinger operators and  give further motivation for them. It turns out that
once chiral boundary conditions have been formulated for the non separable case
further generalization to Schr\"odinger operators with background potential and
to multielectron systems where electrons are allowed to interact,
follow. For the
sake of simplicity and concreteness we shall stick to one electron billiards.
Moreover, to avoid writing complicated formulas, we shall assume that the two dimensional manifold
$M$ is (metrically) cylindrical near its boundary $\partial M$.

It is instructive to formulate the chiral boundary conditions in terms of
quadratic forms, and to compare them with the classical boundary
conditions, Dirichlet and Neumann.
A positive quadratic form, $Q(\varphi)$, on a dense domain,  uniquely
defines a
self-adjoint operator \cite{kato}. The nice thing about quadratic forms is that
the boundary conditions are part of the form and suggest a physical
interpretation. Let
$\langle \cdot \vert \cdot \rangle_{M}$
stands for the scalar product in
$L^2( M)$ and $\langle \cdot \vert \cdot\rangle_{\partial M}$ for the scalar
product on the boundary of $M$.  $ C^\infty ( M)$ is the space of smooth
functions on $M$. The quadratic form
\begin{equation}  Q(\varphi) = \langle \nabla \varphi \vert \nabla \varphi
\rangle_{M} +\lambda  \left\langle \varphi \vert \varphi
\right\rangle_{\partial M}
\label{DN}
\end{equation}
with $\varphi\in C^\infty (M)$ and $0\le\lambda<\infty$,
describes for $\lambda =0$ the Neumann problem  and for
$\lambda \to \infty$ the Dirichlet problem for  the Laplacian $\Delta$.  For
finite
$\lambda$ one has the elastic boundary conditions.  The
Neumann problem says that the boundary term gives no penalty (in energy)  if
there is density on the boundary, while, Dirichlet says that the penalty is
large and so finite energies have zero density on the boundary. It is an
immediate consequence of the quadratic form and the variational principle that
the Dirichlet spectrum  have energies above the Neumann spectrum. $\lambda$
scales like the inverse of a length squared
so that in the absence of a dimensional
parameter, Dirichlet and Neumann are distinguished.

Dirichlet and Neumann  associate a penalty for density at the boundary. Chiral
boundary conditions associate a penalty for a chirality. Since we want edge
states (which have positive chirality) to pay a price and bulk  states (which
have negative chirality) not to affected by the boundary, a
quadratic form which does that in the presence of gauge fields is:
\begin{eqnarray}  Q_c(\varphi) &=&
\langle D \varphi \vert D \varphi \rangle _{M}
+\lambda  \Big\langle \varphi \Big\vert v_+ \varphi
\Big\rangle_{\partial M}
\nonumber \\
v_+&=& \cases{v_y\quad &$if\  v_y>0$;\cr 0 &$otherwise$,}\label{chiral}
\end{eqnarray}
where $\varphi\in C^\infty (M)$, $0\le\lambda<\infty$
and $v_y$ is the operator of (tangential) velocity on the boundary.
Now, in contrast to the Dirichlet-Neumann case discussed above,  $\lambda$
is dimensionless.
To see what  this implies for the boundary conditions we need to go to the
operator and its domain.
The domain of $D^\dagger D$ consists of all smooth functions, such
that
\begin{equation}\label{adjoint} \langle D\varphi \vert D \cdot \rangle _{M} +
\lambda\Big\langle
\varphi \Big\vert v_+ \cdot \Big\rangle_ {\partial M},
\end{equation}
is a $L^2$--bounded linear functional.
Integration by parts in the variable
$x$ leads to
\begin{equation} \langle D^\dagger D\varphi \vert \, \cdot \rangle _{M} +
\Big\langle(D+\lambda v_+)\varphi \Big\vert \,  \cdot \Big\rangle_ {\partial M}.
\end{equation}
For this to define a linear functional, the term on the boundary must vanish
identically for all $\varphi$ in the domain of $D^\dagger D$. If  we write
$\varphi=\varphi_++\varphi_-$, where $\varphi_+$ restricted to  $\partial M$
belongs to the positive spectral subspace of $v_y$ this domain is defined by:
$(d_x+(\lambda-1)v_y)\varphi_+ = 0$ and
$D\varphi_- = 0$.
$\lambda =0$ gives spectral boundary condition for both chiralities.
$\lambda=1$ gives spectral boundary conditions for negative chiralities and
Neumann for positive chiralities. In the separable case this gives the
chiral boundary conditions Eq.~(\ref{chiral}). $\lambda=\infty$ gives the APS
boundary conditions. In principle, one could take $\lambda$ as a
parameter in the theory,  fixed by the sound velocity for the edge
states. $\lambda=1$ is distinguished in tending to maximize the density of the
edge states at the boundary.

The quadratic form is gauge invariant and   non-negative and
 therefore defines a non-negative, gauge invariant, Hamiltonian associated to
kinetic energy:
$H_{L} = D^\dagger {D} \geq 0$.
The Hamiltonian $H_{L}$ is symetric by a direct calculation.

Chiral Schr\"odinger Hamiltonians define a self-adjoint
eigenvalue problem. This is true irrespective of whether the problem is
separable or not; if there is a background scalar potential or not, and even if
one considers a one electron theory or a multielectron Hamiltonian.   However,
only in the separable one particle case, (and slightly more general but still
nongeneric cases), does one have a clean splitting of the eigenspaces of the
Hamiltonian into two pieces:
${\cal H}_e$ and
${\cal H}_b$. In general, an  eigenstate $\varphi$  will have both
a non-zero
$\varphi_+$ and $\varphi_-$ piece, and the spectral subspaces do not split
cleanly. The best one might expect in the non separable case is that
in certain limits eigenstates will have a dichotomy. Namely, either
$\varphi_-$ or
$\varphi_+$ will be small in the limit for every eigenstate. Examination of
simple examples suggests that in the limit of large magnetic fields, $B\to
\infty$, there is such an asymptotic splitting.
Similarly, it would be interesting to formulate a corresponding splitting
principle in the multiparticle Fock space. Both questions are open and
interesting.

{\bf Acknowledgment}

This work is supported in part by the  grants from the Israel Academy of
Sciences, the DFG,  and by the fund for promotion of Research at the Technion.
RS acknowledges the hospitality of ITP at the Technion where part of this work
was done.

\end{document}